\documentclass[aps, prd, 12pt, eqsecnum, tightenlines, notitlepage, nofootinbib, floatfix]{revtex4-1}
\usepackage[bookmarksnumbered=true,bookmarksopen=true]{hyperref}
\usepackage{multirow, slashed, xcolor, amsmath}
\usepackage{tensor}
\usepackage{mathtools}
\usepackage{eucal}
\usepackage{float}
\usepackage{verbatim}
\usepackage{feynmp}
\usepackage{subfig}

\makeatletter
\def\endfmffile{
  \fmfcmd{\p@rcent\space the end.^^J end.^^J endinput;}
  \if@fmfio
    \immediate\closeout\@outfmf
  \fi
  \ifnum\pdfshellescape>\z@
    \immediate\write18{mpost \thefmffile}
  \fi}
\makeatother

\usepackage{multirow}
\DeclareRobustCommand{\vect}[1]{
  \ifcat#1\relax
    \boldsymbol{#1}
  \else
    \mathbf{#1}
  \fi}
  
\newcommand{\de}{\mathrm{d}}
\usepackage{amsmath,amssymb}

\newcommand{\marrow}[5]{%
    \fmfcmd{style_def marrow#1
    expr p = drawarrow subpath (1/4, 3/4) of p shifted 6 #2 withpen pencircle scaled 0.4;
    label.#3(btex #4 etex, point 0.5 of p shifted 6 #2);
    enddef;}
    \fmf{marrow#1,tension=0}{#5}}

\newcommand{\marrowbis}[5]{%
    \fmfcmd{style_def marrow#1
    expr p = drawarrow subpath (0,2/3) of p shifted .5 #2 withpen pencircle scaled 1;
    label.#3(btex #4 etex, point 0 of p shifted 10 #2);
    enddef;}
    \fmf{marrow#1,tension=0}{#5}}
    
\newcommand{\isEquivTo}[1]{\underset{#1}{\sim}}

\usepackage{xcolor}

\begin{document}

\title{Regularization Scheme Dependence of the Counterterms in the Galaxy Bias Expansion}
\author{Samuel Patrone, Adriano Testa and Mark B. Wise}
\affiliation{Walter Burke Institute for Theoretical Physics, California Institute of Technology, Pasadena, CA 91125}

%%%%%%%%%%%%%%%%%%%%%%%%%%%%%%%%%%%%%%%%%%%%%%%%%%%%%%%%
\begin{abstract}
In this paper we explore how different regularization prescriptions affect the counterterms in the renormalization of the galaxy bias expansion. We work in the context of primordial local non-Gaussianity including non-linear gravitational evolution. We carry out the one-loop renormalization of the field $\delta_\rho^2$ (i.e. the square of the matter overdensity field) up to third order in gravitational evolution. Three regularization schemes are considered and their impact on the values of the counterterms is studied. We explicitly verify that the coefficients of the non-boost invariant operators are regularization scheme independent.

\end{abstract}
\maketitle
\clubpenalty = 10000
\widowpenalty = 10000

%%%%%%%%%%%%%%%%%%%%%%%%%%%%%%%%%%%%%%%%%%%%%%%%%%%%%%%%
\section{Introduction}
The galaxy bias expansion (for a review see~\cite{Desjacques:2016bnm}) relates the galaxy overdensity field $\delta_g$ (i.e. the relative fluctuations in the number density of galaxies) to the mass overdensity field $\delta_\rho$. In this paper we don't distinguish between dark matter halos and galaxies.
The composite fields in this expansion can be regulated by a short distance cutoff ($1/\Lambda$). Renormalization renders the galaxy overdensity correlators cutoff independent. This has been studied both for Gaussian primordial curvature fluctuations \cite{PhysRevD.74.103512,McDonald_2009,Assassi_2014} and for non-Gaussian primordial curvature fluctuations~\cite{McDonald_2008,Assassi_2015}. 

In Section~\ref{sec:Framework}, we review the theoretical framework of local non-Gaussianity for primordial curvature fluctuations. The non-linear gravitational evolution of the matter overdensity field is discussed. We finally introduce three regularization schemes for the composite operator\footnote{In this work we use the terms field and operator interchangeably.} $\delta_{\rho}^2$ occurring in the galaxy bias expansion. 

In Section~\ref{sec:renormalization}, we re-examine the renormalization procedure for the composite operator $\delta_{\rho}^2$ in the presence of primordial local non-Gaussianities up to third order in gravitational evolution. We explore the impact of three regularization prescriptions on renormalization. The first reproduces the coefficients and operators found in the literature \cite{Assassi_2015}. The other two prescriptions close under renormalization using the same operators as the first one. However, in the presence of non-linear gravitational evolution, the coefficients of the boost-invariant operators for the three prescriptions differ by quantities which can be written as surface terms in the UV diverging integrals.

\enlargethispage*{.5cm}
In Appendix \ref{appendix:CI}, after introducing a diagrammatic notation, we prove the conformal invariance of the tree-level correlators of curvature fluctuations in the context of local primordial non-Gaussianity. In Appendix \ref{appendix:feyn_rules}, we introduce a diagrammatic notation to compute and estimate the correlators of the galaxy overdensity field and comment on an alternative way of regularizing IR divergences.

This paper concerns the formalism of renormalization and it is only of pedagogical value since the results will not impact the comparison with observations.
%%%%%%%%%%%%%%%%%%%%%%%%%%%%%%%%%%%%%%%%%%%%%%%%%%%%%%%%
\section{Theoretical Framework} 
\label{sec:Framework}
%%%%%%%%%%%%%%%%%%%%%%%%
\subsection{Primordial Curvature Fluctuations and Local non-Gaussianity}
\label{sec:primordial_curvature}
Local non-Gaussianity is the hypothesis that the primordial curvature fluctuations $\delta_{\zeta}$ are local functions of a Gaussian field $\phi_G$ of the form
\begin{equation}
\label{eq:f_expansion}
\delta_\zeta ({\bf x})=\sum_{n=1}^{\infty} f_n:\phi_G^n({\bf x}):\,.
\end{equation}
In the above equation, operators surrounded by colons are normal ordered (i.e. any contraction between themselves in correlators vanishes), $\phi_G$ is a Gaussian field with zero expectation value and the $f_n$'s are constants. In this paper we set $f_1=1$. Demanding the correlators of $\phi_G$ to be invariant under translations, rotations and scale transformations fixes the two-point function of $\phi_G$ to be
\begin{equation}
\label{eq:phipropagator}
\langle {\tilde \phi_G} ({\bf k}_1) {\tilde \phi_G} ({\bf k}_2) \rangle= (2 \pi)^3 \delta({\bf k}_1+{\bf k}_2) \,P_{\phi}(k_1)\,,
\end{equation}
where a tilde denotes a wavevector space quantity, $P_{\phi}(q)=A/q^{3-2\Delta}$ is the tree-level power spectrum of primordial curvature fluctuations ($A$ is a constant), $\Delta$ is the scale dimension of the field $\phi_G$, and $k_i=|{\bf k}_{i}|$. In this paper we will set $\Delta = 0$.

Measurements of the Cosmic Microwave Background (CMB) anisotropy place bounds \cite{Planck:2018vyg} on the local non-Gaussianity parameters, in particular (with the convention $f_1 = 1$) $A \sim  10^{-8}$, $f_2=3/5 f_{NL}^{\rm local} =3/5 (0.8 \pm 5)$, $f_3= (9/25)g_{NL}^{\rm local}$ and $g_{NL}^{\rm local}=(-9.0\pm 7.7)\times 10^4$. It can be seen, by rescaling $\phi_G \to \sqrt{A} \phi_G$ and factoring out a common $\sqrt{A}$, that a highly non-Gaussian theory would have $f_n \sim \mathcal{O}\left[A^{(-n+1)/2}\right]$. Hence, the above CMB bounds already indicate that our universe has nearly-Gaussian primordial curvature fluctuations.

In Appendix \ref{appendix:CI}, after introducing a diagrammatic notation for the computation of the $\delta_\zeta$ $N$-point functions, we prove by induction the conformal invariance of these correlators at tree-level.
%%%%%%%%%%%%%%%%%%%%%%%%
\subsection{Matter Density Perturbations}
\label{sec:matter_correlators}
Even when the primordial curvature fluctuations are Gaussian, non-Gaussianities in the matter overdensity field arise at late times due to the non-linear gravitational evolution.

For simplicity, we shall assume that the all the matter in the universe is in the form of cold dark matter and behaves as a pressureless irrotational fluid. Labelling each fluid element trajectory by its initial position ${\bf x}_0$ (i.e.~its Lagrangian coordinate), its Eulerian coordinate at conformal time $\tau$ is 
\begin{equation}
\label{xofq}
    {\bf x}({\bf x}_0, \tau) = {\bf x}_0 + {\bf s}({\bf x}_0, \tau)
\end{equation}
where
\begin{equation}
{\bf s}({\bf x}_0, \tau) = \int_0^{\tau}d\tau'\, {\bf v}({\bf x}({\bf x}_0, \tau'), \tau')
\end{equation}
is called the displacement vector and ${\bf v}$ is the fluid element velocity.

Using standard gravitational perturbation theory 
\cite{Peebles,Fry:1983cj,Goroff:1986ep} in the Newtonian approximation, the solutions to the Euler and Poisson equations in an expanding universe for the overdensity matter field $\delta_\rho({\bf x}, \tau)$ can be written as
\begin{equation}
    \delta_\rho({\bf x}, \tau) =  D(\tau)\delta_\rho^{(1)}({\bf x}) + D^2(\tau)\delta_\rho^{(2)}({\bf x}) + D^3(\tau)\delta^{(3)}_{\rho}({\bf x})+ \dots\,,
\end{equation}
where $D(\tau)$ is the growth factor, $D(\tau)\delta^{(1)}_\rho({\bf x})$ is the solution to the linearized equations and we kept only the fastest growing modes at each order. We can express $\tilde\delta_\rho^{(n)}$ as
\begin{equation}
\label{deltarhon}
    \tilde{\delta}_\rho^{(n)}({\bf k}) =  \left[\prod_{i=1}^n \int \frac{{\rm d^3}k_i}{(2\pi)^3}\right] (2\pi)^3  \delta_{\rm D}^{(3)}\left({\bf k}-\sum_{i=1}^n  {\bf k}_i\right) F_n({\bf k}_1,\cdots, {\bf k}_n) \tilde\delta_\rho^{(1)}({\bf k}_1)\cdots\tilde\delta_\rho^{(1)}({\bf k}_n)\,,
\end{equation}
where the $F_n$'s are called splitting functions and $F_1=1$. 
The velocity field is described by its divergence $\theta({\bf x},\tau)=\nabla\cdot{\bf v({\bf x},\tau)}$ which, similarly to the field $\delta_\rho({\bf x}, \tau)$, can be written as a perturbative solution of the Euler and Poisson equations as
\begin{equation}
     \theta({\bf x}, \tau) =  - \frac{{\rm d} D(\tau)}{{\rm d} \tau} \left( \theta^{(1)}({\bf x}) + D(\tau)\theta^{(2)}({\bf x}) + \dots\,\right)\,
\end{equation}
where 
\begin{equation}
\label{thetan}
    \tilde{\theta}^{(n)}({\bf k}) =  \left[\prod_{i=1}^n \int \frac{{\rm d^3}k_i}{(2\pi)^3}\right] (2\pi)^3  \delta_{\rm D}^{(3)}\left({\bf k}-\sum_{i=1}^n  {\bf k}_i\right) G_n({\bf k}_1,\cdots, {\bf k}_n) \tilde\delta_\rho^{(1)}({\bf k}_1)\cdots\tilde\delta_\rho^{(1)}({\bf k}_n)\,
\end{equation}
and $G_1=1$.
In this paper, we only need the explicit expressions \cite{Jeong} for $F_2$, $G_2$ and $F_3$, which are
\begin{align}
    F_2({\bf k}_1, {\bf k}_2) = \frac{5}{7} + \frac{1}{2} \frac{{\bf k}_1\cdot{\bf k}_2}{k_1 k_2}\left(\frac{k_1}{k_2}+ \frac{k_2}{k_1}\right)+\frac{2}{7}\frac{({\bf k}_1\cdot{\bf k}_2)^2}{k_1^2 k_2^2}\,,\\
        G_2({\bf k}_1, {\bf k}_2) = \frac{3}{7} + \frac{1}{2} \frac{{\bf k}_1\cdot{\bf k}_2}{k_1 k_2}\left(\frac{k_1}{k_2}+ \frac{k_2}{k_1}\right)+\frac{4}{7}\frac{({\bf k}_1\cdot{\bf k}_2)^2}{k_1^2 k_2^2}\,,\\
F_3({\bf k}_1,{\bf k}_2,{\bf k}_3) = \frac{2 k_{123}^2}{54}\left[\frac{{\bf k}_1 \cdot {\bf k}_{23}}{k_1^2 k^2_{23}}G_2({\bf k}_{2},{\bf k}_{3}) +(2 \textrm{ cyclic}) \right]\nonumber\\
+\frac{7}{54}{\bf k}_{123}\cdot\left[\frac{{\bf k}_{12}}{k^2_{12}}G_2({\bf k}_{1},{\bf k}_{2}) +(2 \textrm{ cyclic}) \right]\nonumber\\
+\frac{7}{54} {\bf k}_{123}\cdot\left[\frac{{\bf k}_{1}}{k^2_{1}}F_2({\bf k}_{2},{\bf k}_{3}) +(2 \textrm{ cyclic}) \right]\,,
\end{align}
where ${\bf k}_{i j l \dots} ={\bf k}_{i} + {\bf k}_{j} +{\bf k}_{l} + \cdots$\,.

The matter overdensity perturbation field at linear order $\tilde\delta^{(1)}_\rho$ is related to the primordial curvature fluctuation $\tilde\delta_\zeta$ by
\begin{equation}\label{eq:M}
\tilde\delta^{(1)}_\rho({\bf k})= M(k) \tilde\delta_\zeta({\bf k})
\end{equation}
where for small wavevectors $M(k)\propto k^2$, and for large wavevectors (below the nonlinear scale) $M(k) \sim 
\sqrt{k}$ \cite{Assassi_2014}.
For the purposes of analytical estimates, we will use the following approximate expression for $M(k)$ 
\begin{equation}\label{eq:M_theta}
    M(k) \simeq \left(\frac{2}{5\Omega_m}\right)\left[\frac{k^2}{H_0^2}\theta(q_0 -k) + \sqrt{\frac{k}{q_0}}\frac{q_0^{2}}{H_0^2}\theta(k -q_0)\right]\,.
\end{equation}
Here $H_0\simeq 70 \,\textrm {km}\,\textrm{s}^{-1}\textrm{Mpc}^{-1}$ is the Hubble constant today, $\Omega_m\simeq 0.3$ is the fraction of matter energy density, and $q_0\simeq 0.015 \,\textrm {Mpc}^{-1}$.

Matter overdensity correlators are computed using the Effective Field Theory of Large Scale Structure (EFT-of-LSS) \cite{Carrasco:2012cv,Baumann_2012}.
We work only to leading order in this theory.

%%%%%%%%%%%%%%%%%%%%%%%%
\subsection{Galaxy Density Perturbations and Regularization Schemes}
\label{subsec:galaxy_bias_expansion}
\label{sec:galaxy_correlators}
As mentioned in the Introduction, the galaxy overdensity field $\delta_g$ can be written as a biased tracer of the underlying matter overdensity field $\delta_\rho$
\begin{equation}
\begin{split}
\label{biasexp}
\delta_g({\bf x}) = b_1 \delta_\rho({\bf x}) + \frac{b_2}{2} \delta^2_\rho({\bf x}) + \ldots\,.
\end{split}
\end{equation} 
In the equation above we omitted operators that are outside the scope of this paper as well as the terms that set the expectation value of $\delta_g$ to zero. The general form of the additional terms represented by the ellipses in the bias expansion above is known \cite{Desjacques:2016bnm} (even for different forms of the power spectrum) and it could include other operators that do not appear as counterterms \cite{Baldauf:2016sjb}.

The composite operators in Eq.~\eqref{biasexp} need to be regularized and renormalized. In this paper we will focus on the operator $\delta_\rho^2$ as an illustration of the dependence of the counterterms on the regularization scheme. What we conclude could be generalized to other composite operators in the galaxy bias expansion. Generically, regularization is achieved by introducing a wavevector cutoff $\Lambda$ in divergent integrals. Here we present three possible choices for the cutoff regulator.

The first regularization scheme we consider consists in cutting off the large wavevector component of the linearized solution $\delta_\rho^{(1)}$ of the gravitational evolution equations, i.e. performing the following replacement
\begin{equation}
\delta_\rho^{(1)}({\bf q})\to \delta_\rho^{(1)}({\bf q}) \theta(\Lambda-q)\,.
\end{equation}
As we will discuss in Sec.~\ref{sec:renormalization} this prescription reproduces the results previously found in \cite{Assassi_2015} after assuming the UV asymptotic behavior for $M$ given in Eq.~\eqref{eq:M_theta}. This prescription has the advantage that, had we worked at higher order in the EFT-of-LSS, it would have regulated both the correlators of $\delta_\rho$ and of the composite operator $\delta_\rho^2$.

Since we are working to lowest order in the EFT-of-LSS we introduce a second and a third renormalization prescriptions for $\delta_\rho^2$ that do not explicitly cut off large wavevectors in the non-linear gravitational evolution equations. 

One choice is to cut off the large wavevectors of the \emph{full} solution $\delta_\rho$ of the gravitational equations, i.e.
\begin{equation}
\label{eq:secondpres}
\delta_\rho({\bf q})\to \delta_\rho({\bf q}) \theta(\Lambda-q)\,,
\end{equation}
which implies
\begin{equation}
\tilde\delta^2_\rho({\bf k})\to \int \frac{{\rm d}^3 q}{(2\pi)^3}\tilde\delta_\rho({\bf q})\tilde\delta_\rho({\bf k-q})\theta(\Lambda-q)\theta(\Lambda-|{\bf k-q}|)\,.
\end{equation}
The third prescription is obtained from the second one by expanding for $k\ll q$ one of the two thetas in the convolution integral of $\delta_\rho^2$ and keeping only the leading term. 
\begin{align}
\label{eq:regdelta2}
\tilde\delta^2_\rho({\bf k})\to& \int \frac{{\rm d}^3 q}{(2\pi)^3}\tilde\delta_\rho({\bf q})\tilde\delta_\rho({\bf k-q})\theta(\Lambda-q)\nonumber\\=& \int \frac{{\rm d}^3 q}{(2\pi)^3}\tilde\delta_\rho({\bf q})\tilde\delta_\rho({\bf k-q})\left[\frac{\theta(\Lambda-q)+\theta(\Lambda-|{\bf k-q}|)}{2}\right]\,.
\end{align}
In the square bracket of the above equation we explicitly symmetrized the regularization kernel, thus rendering the expression manifestly symmetric in the two $\tilde\delta_\rho$'s. 
This last prescription leads to a spherically symmetric integration region (hence simplifying the integrals).

We find that all three regulators give the same non-boost invariant terms\footnote{In this paper we refer to operators that diverge as a single wave-vector vanishes as non-boost invariant.} and that they can be removed by expressing $\phi_G$ in Lagrangian coordinates as expected from the results of \cite{Assassi_2015, Mirbabayi:2014zca}. After assuming a UV asymptotic behavior for $M$ (e.g. the one in Eq.~\eqref{eq:M_theta}), the second and third regulators give the same coefficients for the boost invariant terms which, however, differ from the ones in the first prescription. Hence, in the following, we will only discuss the first and the third regularization schemes.

We will deal with infrared divergences that arise in the computation of correlators of the galaxy overdensity field by expanding the integrands around $q=\infty$ and retaining only the terms that contribute to the UV divergence. These terms are infrared safe.

%%%%%%%%%%%%%%%%%%%%%%%%%%%%%%%%%%%%%%%%%%%%%%%%%%%%%%%%

\section{Renormalization of the operator $\delta_{\rho}^2$ at one loop}
\label{sec:renormalization}
The correlators of $\delta_g$ are sensitive to the physics of large wavevectors where perturbation theory is no longer valid. To make analytic predictions using a perturbative approach, composite operators in the galaxy bias expansion need to be regularized and renormalized adding counterterms that remove the sensitivity to the physics of large wavevectors.
Following \cite{PhysRevD.74.103512}, we only keep the fastest growing modes in the computation of the counterterms.

Note that the composite operator $\delta^2_\rho$ can ultimately be written as a convolution of fields $\phi_G$ at different wavevectors. Therefore, we find the one-loop counterterms by contracting two such fields, introducing a UV regulator $\Lambda$ (as described in Sec.~\ref{sec:galaxy_correlators}), and by selecting the parts that diverge as $\Lambda$ goes to infinity.

Even though all the amplitudes presented in this paper can be obtained applying Wick's theorem, such operation can be tedious and cumbersome. Therefore, we will use a diagrammatic formalism to graphically keep track of the different contributions in the renormalization of the operator $\delta_\rho^2$ (see Appendix~\ref{appendix:feyn_rules} for details). Diagrams with $N$ cross vertices represent contributions to the $N$-point galaxy overdensity field correlator $\langle \tilde\delta_g({\bf k}_1) \dots\tilde\delta_g({\bf k}_N)\rangle$.

In this section, we discuss the one-loop renormalization of the quadratic operator $\delta_\rho^2$ for local primordial non-Gaussianity.
At the end of the section, we give the full renormalized expression of the operator $\delta_\rho^2$ up to third order in gravitational evolution in the first and the third renormalization schemes introduced above. Using the first regularization prescription, we reproduce the results found in \cite{Assassi_2015} after assuming the UV asymptotic behavior for $M$ of Eq.~\eqref{eq:M_theta}.
\subsection{First Order in Gravitational Evolution}

\begin{figure}[t]
\vspace{20pt}
\input{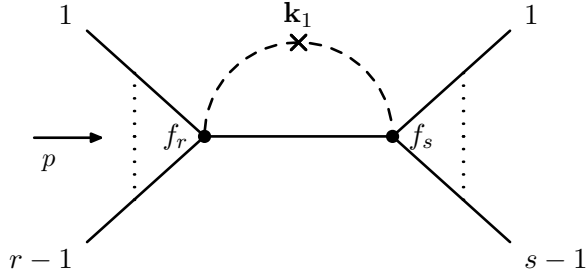}
\vspace{20pt}
\caption{Generic one-loop divergent diagrams at first order in gravitational evolution.}
\label{fig:LO_sigma}
\end{figure}

In this subsection, we work at linear order in gravitational evolution, i.e.~we set all the $F_n$'s and $G_n$'s (for $n>1$) to zero. We compute the counterterms needed to renormalize the operator $\delta_{\rho}^2$ at one loop in the presence of primordial non-Gaussian perturbations of the form of Eq.~\eqref{eq:f_expansion}. In Fig.~\ref{fig:LO_sigma}, we show all the (amputated) one-loop divergent diagrams with a single insertion of $\delta_{\rho}^2$. These diagrams are obtained by contracting a single $\phi_G$ from each $\delta_\rho$ in $\delta_{\rho}^2$. 

Here, the first regularization prescription gives
\begin{equation}\label{eq:sigma12}
      (2-\delta_{rs}) r s f_r f_s \int \frac{\de^3 q}{(2 \pi)^3} M(q) M(|\mathbf{k}_1-\mathbf{q}|)P_{\phi}(|\mathbf{p}+\mathbf{q}|) \theta(\Lambda - q)\theta(\Lambda -|{\bf k}_1-{\bf q}|)\,
\end{equation}
where $\mathbf{p}$ is the total wavevector entering the $f_r$ vertex. We evaluate this kind of integral in cylindrical coordinates where the product of the two thetas can naturally be embedded in the integration measure. 

The third regularization prescription gives
\begin{equation}\label{eq:sigma3}
      (2-\delta_{rs}) r s f_r f_s \sigma^2({\bf k}_1,{\bf p}; \Lambda)
\end{equation}
where
\begin{equation}\label{eq:sigmafull}
\setlength{\jot}{10pt}
   \sigma^2({\bf k}_1,{\bf p};\Lambda)= \int \frac{\de^3 q}{(2 \pi)^3} M(q) M(|\mathbf{k}_1-\mathbf{q}|)P_{\phi}(|\mathbf{p}+\mathbf{q}|)\left[\frac{\theta(\Lambda-q)+\theta(\Lambda-|{\bf k}_1-{\bf q}|)}{2}\right] \,.
\end{equation}
We embed the thetas in the integration measure and we expand the remaining part of the integrand around $q=\infty$, using the large wavevector asymptotic expression of Eq.~\eqref{eq:M_theta} for $M$. We keep only the powers that give rise to UV divergent terms (i.e. all the terms up to $q^{-2}$). Note that terms in the integrand suppressed by powers of k/q give rise to finite contributions that are dropped. These terms do not contain IR or collinear divergences. After this procedure the divergent parts of the integral in Eq.~\eqref{eq:sigma12} are
\begin{equation}
\sigma^2_{\rm asy} \equiv \dfrac{4}{25}\dfrac{Aq^3_0}{\Omega^2_mH_0^4} \dfrac{ \Lambda}{2\pi^2}\,.
\end{equation}

Due to the spherical symmetry of the integration region of Eq.~\eqref{eq:sigmafull}, for the third prescription an alternative procedure can be followed. Expanding the integrand for $k_1, p \ll q$ and performing the angular integral ${\rm d \Omega}_q$, the linearly divergent part of Eq.~\eqref{eq:sigmafull} is
\begin{equation}
 \sigma^2\equiv \sigma^2(0,0; \Lambda)=\int\frac{\de^3 q}{(2 \pi)^3} M^2(q)P(q) \theta(\Lambda - q)\,.
\end{equation}
$\sigma^2$ coincides with $\sigma^2_{\rm asy}$ when using the UV asymptotic expression for $M$ of Eq.~\eqref{eq:M_theta}.

We therefore introduce the following position space counterterms to make correlators involving the operator $\tilde\delta_\rho^2$ finite as $\Lambda$ goes to infinity
\begin{equation}
\label{sigmacounterterm}
    c_{r-1,s-1}(\Lambda):\phi_G^{r-1}({\bf x})::\phi_G^{s-1}({\bf x}):\,
\end{equation}
where 
\begin{equation}\label{sigmacountertermI}
c_{r-1,s-1}(\Lambda)=- r s  f_r f_s \sigma_{\rm asy}^2\,
\end{equation}
and $\sigma^2$ can be used in place of $\sigma^2_{\rm asy}$. 
At one loop the two normal orderings in Eq.\,\eqref{sigmacounterterm} can be merged to a single overall one, and consequently the counterterms take the form
\begin{equation}\label{eq:sigma_coeff}
c_{n}(\Lambda):\phi_G^n({\bf x}):\,= \left(\sum_{i=0}^n c_{i,n-i}(\Lambda)\right):\phi_G^n({\bf x}):\,.
\end{equation}
As mentioned above the first and the third prescriptions give the same counterterms which up to second order in the field $\phi_G$ are
\begin{eqnarray}
   -\sigma_{\rm asy}^2\left[4 f_2 \, \phi_G({\bf x}) +(6 f_3 + 4f_2^2)\,:\phi_G^2({\bf x}):\right]\,.
\end{eqnarray}

\begin{figure}[t]
\input{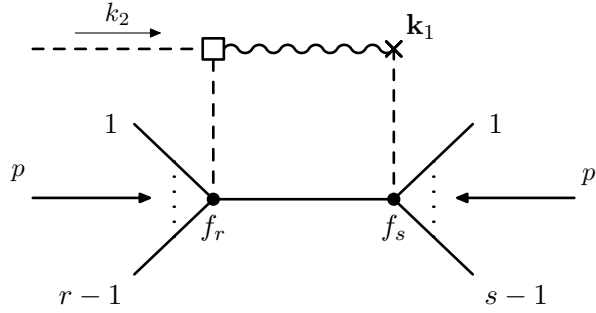}
\vspace{0.5cm}
\caption{Diagrams with a single $F_2$ vertex used to determine the counterterm for $\delta_{\rho}^2$ linear in $\delta_{\rho}$.}
\label{fig:boxgen}
\end{figure}

\subsection{Second Order in Gravitational Evolution}

We now study the effect of non-linear gravitational evolution on renormalization for non-Gaussian primordial fluctuations of the local type. 
In this case the renormalization of $\delta_\rho^2$ will generate counterterms linear in $\delta_\rho^{(1)}$ (which will reassemble to $\delta_\rho$ when higher order terms are included).

We begin by considering Gaussian primordial fluctuations (see Fig.~\ref{fig:boxgen} choosing $r=1$ and $s=1$) and we evaluate its divergent coefficient in the first and third regularization prescriptions described in Sec.~\ref{subsec:galaxy_bias_expansion}.

The first prescription gives 
\begin{equation}
\label{eq:gauss_box_1}
   4 \int\frac{{\rm d}^3 q}{(2\pi)^3}  F_2({\bf q},-{\bf k}_1) M^2(q) P_{\phi}(q)\theta(\Lambda - q) \,,
\end{equation}
where we dropped thetas of the type $\theta(\Lambda - k_i)$ since $k_i<\Lambda$\,.
Expanding for small $k_1/q$ and performing the angular integral ${\rm d \Omega}_q$, the linearly divergent part of the above equation is
\begin{equation}\label{eq:3.11}
\frac{68}{21} \int\frac{{\rm d}q}{2\pi^2}   q^2 M^2(q)P_{\phi}(q)\theta(\Lambda-q)=\frac{68}{21}\sigma^2 \simeq \frac{68}{21}\sigma^2_{\rm asy}\,,
\end{equation}
where in the last step we assumed the UV asymptotic behavior of Eq.\eqref{eq:M_theta} for $M$.
This result reproduces the counterterms already found in the literature \cite{Assassi_2014}.

\begin{figure}[t]
\vspace{20pt}
\input{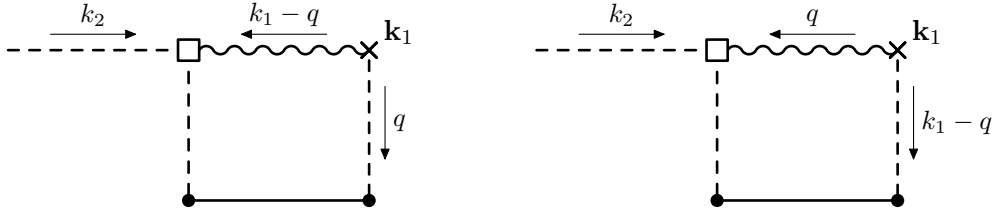}
\vspace{20pt}
\caption{Two routings of the loop diagram with a single $F_2$ vertex used to determine the counterterm for $\delta_{\rho}^2$ linear in $\delta_{\rho}$ with Gaussian primordial fluctuations.}
\label{fig:Gauss_F2}
\end{figure}

For the third prescription, $\tilde\delta_\rho^2$ at this order is
\begin{equation}
\label{eq:routingamb}
   \int \frac{{\rm d}^3 q}{(2\pi)^3} \left[\tilde\delta_\rho^{(2)}({\bf q})\tilde\delta_\rho^{(1)}({\bf k}_1-{\bf q})+\tilde\delta_\rho^{(1)}({\bf q})\tilde\delta_\rho^{(2)}({\bf k}_1-{\bf q})\right]\theta(\Lambda-q)\,.
\end{equation}
Contracting one of the $\tilde\phi_G$ fields in $\tilde\delta_\rho^{(2)}$ with the $\tilde\phi_G$ field in $\tilde\delta_\rho^{(1)}$ gives an operator proportional to $\tilde\delta_\rho^{(1)}({\bf k}_1)$ with (divergent) coefficient
\begin{equation}
\label{eq:gauss_box}
   2 \int\frac{{\rm d}^3 q}{(2\pi)^3} \left[ F_2(-{\bf q},{\bf k}_1) M^2(q) P_{\phi}(q)+ F_2(-{\bf k}_1+{\bf q},{\bf k}_1) M^2(|{\bf k}_1-{\bf q}|) P_{\phi}(|{\bf k}_1-{\bf q}|)\right]\theta(\Lambda-q) \,.
\end{equation}
In diagrammatic notation, the two terms above are represented by the (amputated) diagrams of Fig.~\ref{fig:Gauss_F2} and correspond to two ways of \textit{routing} the loop wavevector.
We observe that the two terms in the square bracket can be made equal to each other with the change of variable ${\bf q}\rightarrow{\bf k}_1-{\bf q}$, however the \textit{integrands} are not equal because of the presence of the theta function. Thus, we expect the above integrals to differ by a surface term which will impact the explicit form of the counterterms.

Expanding for small $k_1/q$ and performing the angular integral ${\rm d \Omega}_q$, the linearly divergent parts of Eq.\,\eqref{eq:gauss_box} are
\begin{equation}
\label{eq:gauss_box_expanded}
     \int\frac{{\rm d}q}{2\pi^2} \left\{\frac{34}{21} q^2M^2(q)P_{\phi}(q) + \frac{34}{21} q^2 M^2(q)P_{\phi}(q) - \frac{1}{3}\frac{{\rm d}}{{\rm d}q} \big(q^3 M^2(q)P_{\phi}(q)\big)\right\} \theta(\Lambda-q)=\frac{68}{21}\sigma^2-\frac{1}{3}\rho^2\,
\end{equation}
where
\begin{eqnarray}
\label{eq:rho}\rho^2&&=\int\frac{{\rm d }q }{2\pi^2}\left[\frac{{\rm d}}{{\rm d}q} \big(q^3 M^2(q)P_{\phi}(q)\big)\right] \theta(\Lambda-q)\,.
\end{eqnarray}
On the left hand side of Eq.~\eqref{eq:gauss_box_expanded}, the first term corresponds to the first routing while the other two terms correspond to the second routing. Notice that $\rho^2$ coincides with $\sigma^2_{\rm asy}$ when using the UV asymptotic expression for $M$ of Eq.~\eqref{eq:M_theta}. As anticipated, the contributions from the two routings differ by a total derivative term. Since this term is linearly divergent, it will contribute to the value of the counterterm. 
We observe that the first term (bulk) on the right hand side of Eq.~\eqref{eq:gauss_box_expanded} is the same as the counterterm for the first prescription of Eq.~\eqref{eq:gauss_box_1}. In loops containing box vertices (representing non-linear gravitational evolution), the third prescription - in diagrammatic formalism - entails symmetrizing the contribution of a diagram over the two possible routings in analogy to Fig.~\ref{fig:Gauss_F2}.

Non-Gaussian primordial fluctuations generate counterterms proportional to the operators $\delta_\rho^{(1)}\phi_G^{n}$. To determine their divergent coefficients we evaluate the diagrams in Fig.~\ref{fig:boxgen} in the regularization prescriptions described above. For each choice of $r$ and $s$ (s.t. $r+s = n$) we need to sum over two diagrams. Letting $\bf p$ and ${\bf p}'$ be the sum of the wavevectors entering the $f_r$ and $f_s$ vertices, respectively, the total amplitude is
\begin{align}
\label{eq:boxgen}
\begin{split}
   2 \, r s f_r f_s \int\frac{{\rm d}^3 q}{(2\pi)^3} &F_2({\bf k}_1+{\bf k}_2-{\bf q},-{\bf k}_2) M(q)M(|{\bf k}_1+{\bf k}_2-{\bf q}|)\left[P_{\phi}(|{\bf p}+{\bf q}|)+P_{\phi}(|{\bf p'}+{\bf q}|)\right] \mathcal{R}_i \,
\end{split}
\end{align}
where the $\mathcal{R}_i$ is the regularization kernel in scheme $i$ and
\begin{eqnarray}
\mathcal{R}_1&=&\theta(\Lambda-q) \theta(\Lambda-|{\bf k}_1+{\bf k}_2-{\bf q}|)\,,\\[10pt]
\mathcal{R}_3&=&\frac{\theta(\Lambda-q) +\theta(\Lambda-|{\bf k}_1-{\bf q}|)}{2}\label{eq:R3}\,.
\end{eqnarray}
Proceeding as before, we obtain for the divergent parts
\begin{eqnarray}
\label{eq:boxgen_res_1}
 && 2\, r s f_r f_s \left(\frac{68}{21}+\frac{{\bf k}_2\cdot({\bf p} + {\bf p'})}{ k_2^2}\right)\sigma^2_{\rm asy}\,,\\
\label{eq:boxgen_res_3}
    && 2\, r s f_r f_s \left[\left(\frac{68}{21}+\frac{{\bf k}_2\cdot({\bf p} + {\bf p'})}{ k_2^2} \right) \sigma^2 - \frac{1}{3} \rho^2\right] \,.
\end{eqnarray} 
We observe that within each regularization prescription presented in this paper, the Gaussian and non-Gaussian coefficients of the $k$-independent parts coincide \cite{Assassi_2015} up to a factor of $2 r s f_r f_s $ (compare Eqs.~\eqref{eq:3.11} and~\eqref{eq:gauss_box_expanded} with Eqs.~\eqref{eq:boxgen_res_1} and~\eqref{eq:boxgen_res_3}). However, as expected \cite{Mirbabayi:2014zca}, in the non-Gaussian case non-boost invariant terms proportional to ${\bf k}_2\cdot({\bf p} + {\bf p}')/k_2^2$ appear. These shift the argument of the field $\phi_G$ from its Eulerian coordinate ${\bf x}$ to its initial Lagrangian position ${\bf x}_0$ (see Eq.~\eqref{eq:x0shift}) and are the same in the three regularization schemes studied in this paper.

Finally we notice that the two routings in the third prescription generate equal and opposite surface terms that are non-boost invariant. For example, for the counterterm proportional to $\delta_\rho^{(1)} \phi_G$ we get
\begin{align}
\label{b2f2F2expansion}
\begin{split}
     4 f_2\int\frac{{\rm d}q}{2\pi^2}  \bigg\{ &\left(\frac{34}{21} +\frac{{\bf k}_2\cdot{\bf k}_3}{2 k_2^2} \right)q^2 M^2(q)P_{\phi}(q) -
   \left(\frac{{\bf k}_2\cdot{\bf k}_3}{6 k_2^2} + \frac{1}{3}\right)\frac{{\rm d}}{{\rm d}q} \big(q^3 M^2(q)P_{\phi}(q)\big)\\
   + &\left(\frac{34}{21}+\frac{{\bf k}_2\cdot{\bf k}_3}{2 k_2^2}\right) q^2M^2(q)P_{\phi}(q)
+\frac{{\bf k}_2\cdot{\bf k}_3}{6 k_2^2}\frac{{\rm d}}{{\rm d}q} \big(q^3M^2(q)P_{\phi}(q)\big)\bigg\} \theta(\Lambda-q)
\end{split}
\end{align}
where, in this case ${\bf p}={\bf k}_3$ and ${\bf p}'=0$, and the two lines correspond to the two possible routings of the loop wavevector. Again we note that the non-boost invariant surface terms proportional to $({\bf k}_2\cdot{\bf k}_3)/k_2^2$ cancel when considering both routings.

\subsection{Third Order in Gravitational Evolution}
\begin{figure}[t]
\vspace{20pt}
\input{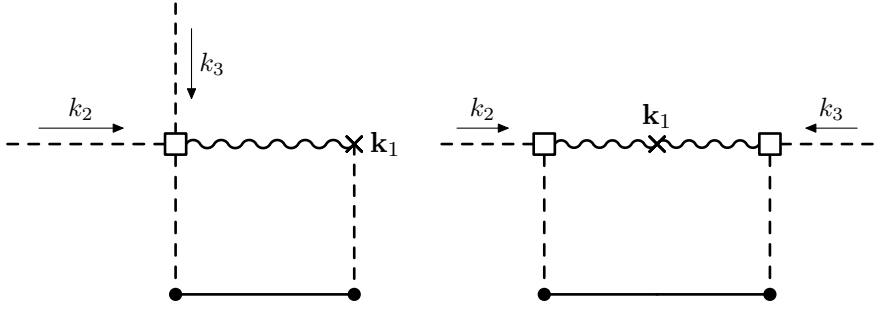}
\vspace{20pt}
\caption{Diagrams with one $F_3$ and two $F_2$'s used to determine the counterterms for $\delta_{\rho}^2$ quadratic in $\delta_{\rho}$ with Gaussian primordial fluctuations.}
\label{fig:3order}
\end{figure} 
We now consider the one-loop counterterms at second order in the field $\delta^{(1)}_\rho$. We have two different diagram topologies (the Gaussian ones are shown in Fig.~\ref{fig:3order}).
The amplitudes for the two topologies at arbitrary order in primordial local non-Gaussianity are
\begin{equation}
\label{F3diag}
\begin{split}
  12 \, r s f_r f_s \int\frac{{\rm d}^3 q}{(2\pi)^3}
   &F_3({\bf k}_1+{\bf k}_2+{\bf k}_3-{\bf q}, -{\bf k}_2, -{\bf k}_3)M(q) M(|{\bf k}_1+{\bf k}_2+{\bf k}_3-{\bf q}|)\\ & \left[P_{\phi}(|{\bf q}+{\bf p}|)+P_{\phi}(|{\bf q}+{\bf p}'|)\right]\mathcal{R}^{(a)}_i\,,
\end{split}
\end{equation}
and
\begin{equation}
\label{F2F2diag}
\begin{split}
   16 \, r s f_r f_s \int\frac{{\rm d}^3 q}{(2\pi)^3}
   &F_2({\bf q}+{\bf k}_2, -{\bf k}_2) F_2({\bf k}_1+{\bf k}_3-{\bf q}, -{\bf k}_3)
   M(|{\bf q}+{\bf k}_2|)M(|{\bf k}_1+{\bf k}_3-{\bf q}|)\\
   &\left[P_{\phi}(|{\bf q} +{\bf p} +{\bf k}_2|) +P_{\phi}(|{\bf q} +{\bf p}' +{\bf k}_2|) \right]\, \mathcal{R}^{(b)}_i \,.
\end{split}
\end{equation}
In the equations above,
\begin{eqnarray}
\mathcal{R}^{(a)}_1&=&\theta(\Lambda-q) \theta(\Lambda-|{\bf k}_1+{\bf k}_2+{\bf k}_3-{\bf q}|)\,,\\
\mathcal{R}^{(b)}_1&=&\theta(\Lambda-|{\bf q}+{\bf k}_2|) \theta(\Lambda-|{\bf k}_1+{\bf k}_3-{\bf q}|)\,,
\end{eqnarray}
and $\mathcal{R}^{(a)}_3=\mathcal{R}^{(b)}_3$ are given in Eq.~\eqref{eq:R3}.

We now give the full renormalized expression of the operator $\delta_\rho^2$ up to third order in the gravitational evolution for local primordial non-Gaussianity. For completeness we reintroduce the subtraction of the vacuum expectation value $\langle\delta^2_\rho\rangle$.

For the first prescription we have
\begin{equation}
\label{eq:delta2renorm2orderfull_1}
\begin{split}
\delta^2_\rho({\bf x})\big|_R=\delta^2_\rho({\bf x}) -& \, \sigma^2_{\rm asy}\mathcal{S}[\phi_G]\left[1 + \frac{68}{21} \delta_\rho({\bf x})+ \frac{2624}{735} \delta^2_\rho({\bf x})+ \frac{254}{2205} \left(\frac{\partial_i\partial_j}{\nabla^2}\delta_\rho({\bf x})\right)\left(\frac{\partial_i\partial_j}{\nabla^2}\delta_\rho({\bf x})\right)\right]\,,
\end{split}
\end{equation}
which reproduces the results of \cite{Assassi_2015} after using the UV asymptotic expression $\sigma_{\rm asy}^2$ in place of $\sigma^2$ for the Gaussian cases.

When using the third prescription we obtain
\begin{equation}
\label{eq:delta2renorm2orderfull_3}
\begin{split}
\delta^2_\rho({\bf x})\big|_R=\delta^2_\rho({\bf x}) -\mathcal{S}[\phi_G]&\left[ \sigma^2  + \left(\frac{68}{21}  \sigma^2 - \frac{1}{3} \rho^2 \right) \delta_\rho({\bf x})+\right.\\
&+ \left(\frac{2624}{735}  \sigma^2 - \frac{73}{105} \rho^2+ \frac{1}{30} \gamma^2\right) \delta^2_\rho({\bf x})+\\
&\left.+ \left(\frac{254}{2205}  \sigma^2 - \frac{16}{105} \rho^2+ \frac{1}{15} \gamma^2\right) \left(\frac{\partial_i\partial_j}{\nabla^2}\delta_\rho({\bf x})\right)\left(\frac{\partial_i\partial_j}{\nabla^2}\delta_\rho({\bf x})\right)\right]\,,
\end{split}
\end{equation}
where
\begin{equation}
\label{gammaint}
\gamma^2=\int\frac{{\rm d}q }{2\pi^2}\left\{\frac{{\rm d}}{{\rm d}q}\left[q^4 \frac{{\rm d}}{{\rm d}q}\big(M^2(q)P_{\phi}(q)\big) \right]\right\}\,\theta(\Lambda-q).
\end{equation}

In the above equations 
\begin{equation}\label{eq:S}
\mathcal{S}[\phi_G] = 1+4f_2\phi_G({\bf x}_0) +(6 f_3 + 4f_2^2):\phi_G^2({\bf x}_0):+\cdots \,
\end{equation}
where, at this order in the gravitational evolution,
\begin{align}\label{eq:x0shift}
\phi_G({\bf x}_0) = \phi_G({\bf x})&+\left(\frac{\partial_\ell}{\nabla^2}\delta_\rho({\bf x})\right)\partial_\ell\phi_G({\bf x}) -\frac{1}{2}\frac{\partial_\ell}{\nabla^2}\left\{\delta_\rho^2({\bf x})- \left(\frac{\partial_i \partial_j}{\nabla^2} \delta_\rho ({\bf x})\right)^2\right\} \partial_\ell\phi_G({\bf x})\nonumber\\
&+\frac{1}{2}\left(\frac{\partial_i }{\nabla^2} \delta_\rho ({\bf x})\right)\left(\frac{\partial_j }{\nabla^2} \delta_\rho ({\bf x})\right)\partial_i\partial_j \phi_G({\bf x})\,
\end{align}
and the ellipses in Eq.~\eqref{eq:S} can be deduced from Eq.~\eqref{sigmacountertermI} and Eq.~\eqref{eq:sigma_coeff}.
In Eq.\eqref{eq:S} the one corresponds to the Gaussian case and the remaining terms arise from primordial non-Gaussianity. The relationship between the counterterms in the Gaussian and non-Gaussian cases was first noted in \cite{Assassi_2015}.

%%%%%%%%%%%%%%%%%%%%%%%%%%%%%%%%%%%%%%%%%%%%%%%%%%%%%%%%
\section{Concluding Remarks}
Composite operators in the bias expansion for the galaxy overdensity field are defined with a large wavevector cutoff ($\Lambda$) and are renormalized to remove the dependence of galaxy overdensity correlators on the wavevector cutoff. In this paper we explored the regularization scheme dependence of the counterterms in the Galaxy bias expansion. We showed by explicit computation how different regularization prescriptions affect the coefficients of the counterterms. As expected the coefficients of the non-boost invariant operators coincide for the regularization schemes explored in this paper and they rearrange to shift the argument of the field $\phi_G$ from Eulerian to Lagrangian coordinates. On the other hand, the boost-invariant terms are dependent on the regularization scheme. Our calculations illustrate the power of the general methods developed in Refs.~\cite{Mirbabayi:2014zca, Assassi_2015}.

%%%%%%%%%%%%%%%%%%%%%%%%%%%%%%%%%%%%%%%%%%%%%%%%%%%%%%%%
\pagebreak
\acknowledgments 
We thank the referee who made important comments on the first version of this paper. This material is based upon work supported by the U.S. Department of Energy, Office of Science, Office of High Energy Physics, under Award Number DE-SC0011632.
We are also grateful for the support provided by the Walter Burke Institute for Theoretical Physics. 

%%%%%%%%%%%%%%%%%%%%%%%%%%%%%%%%%%%%%%%%%%%%%%%%%%%%%%%%

%%%%%%%%%%%%%%%%%%%%%%%%%%%%%%%%%%%%%%%%%%%%%%%%%%%%%%%%
%%%%%%%%%%%%%%%%%%%%%%%%%%%%%%%%%%%%%%%%%%%%%%%%%%%%%%%%
\appendix
\pagebreak
\section{Conformal Invariance of $\delta_\zeta$ correlators at tree-level}
\label{appendix:CI}

In this Appendix, we prove that in local non-Gaussianity all the tree-level correlators of the curvature fluctuations $\delta_\zeta$ are conformally invariant. 

We start by introducing a diagrammatic notation for the $N$-point correlators $P^{(N)}$ defined as
\begin{equation}
\langle\tilde\delta_\zeta(\mathbf{k}_1)\dots\tilde\delta_\zeta(\mathbf{k}_N)\rangle=(2\pi)^3\delta(\mathbf{k}_1+\dots+\mathbf{k}_N)P^{(N)}(\mathbf{k}_1,\dots,\mathbf{k}_N)\,.
\end{equation}
We separate the sum of all the tree-level contributions ($T^{(N)}$) from the sum of the loop contributions using the following notation
\begin{equation}
P^{(N)}(\mathbf{k}_1,\dots,\mathbf{k}_N) = T^{(N)}(\mathbf{k}_1,\dots,\mathbf{k}_N) + \mathrm{loops}\,,
\end{equation}
and we shall refer to $T^{(N)}$ as the tree-level correlator in wavevector space.

\begin{figure}[t]
  \centering
  \input{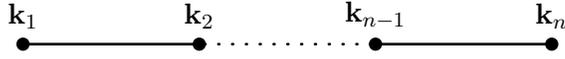}
  \caption{Generic chain diagram.} \label{fig:chaindiag}
\end{figure}

Using $\delta_\zeta$ as in Eq.~\eqref{eq:f_expansion}, we first consider the case where only $f_1$ and $f_2$ are non-zero. Then, it is straightforward to see that the tree-level N-point correlator is
\begin{equation}\label{f2linediag}
  T^{(N)}(\mathbf{k}_1, \dots, \mathbf{k}_N) = 2^{N-2} A^{N-1}f_1^2 f_2^{N-2} \,\frac{1}{2}\sum_{ \mathcal{P}(\mathbf{k}_1 \dots \mathbf{k}_N)}\frac{1}{k_1^3}\frac{1}{|\mathbf{k}_1 + \mathbf{k}_2|^3}\cdots \frac{1}{|\mathbf{k}_1 + \dots + \mathbf{k}_{N-1}|^3} \,, 
\end{equation}
where the sum is over all the permutations $\mathcal{P}(\mathbf{k}_1 \dots \mathbf{k}_N)$ of the wavevectors.
The diagram in Figure \ref{fig:chaindiag} represents the identity permutation contribution to $T^{(N)}$ above.

More generally, contributions to the $N$-point function can be represented by diagrams with $N$ vertices (dots) connected by solid lines.

To compute the connected tree-level $N$-point function of $\tilde{\delta}_\zeta$, we first draw all the connected tree diagrams with $N$ vertices, i.e. different topologies. For each diagram we arbitrarily assign the wavevectors $\mathbf{k}_1, \mathbf{k}_2, \ldots \mathbf{k}_N$ to the various vertices.

\begin{figure}[t]
\centering
\input{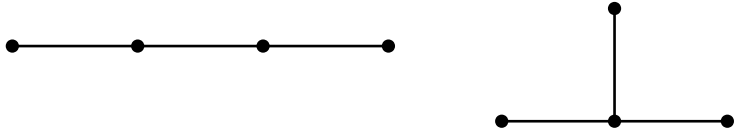}
\vspace{0.5cm}
\caption{Contributions to $T^{(4)}$.}
\label{fig:T4diag}
\end{figure}

We then use the following rules to compute the contribution to $T^{(N)}$ of each diagram. 
\begin{itemize}
    \item Every vertex with $n$ lines emerging from it contributes a factor of $f_n$.
    \item Every line between two vertices contributes a factor $ \dfrac{A}{|\sum_i \mathbf{k}_i|^3}$ where the sum is over the wavevectors that precede and include either of the two vertices along the tree (these two choices are equivalent due to overall wavevector conservation).
    \item Every diagram has a combinatorial factor that is given by the number of ways in which the lines from the various vertices can be joined together.
    \item Every diagram has a symmetry factor that is given by the inverse of the number of symmetries the diagram has. For example, the diagram in Fig.~\ref{fig:chaindiag} has a symmetry factor of $1/2$ because of the reflection symmetry of the chain around its center while the second diagram in Fig.\,\ref{fig:T4diag} has a symmetry factor of $1/3!$.
\end{itemize}

Finally, we sum over the $N!$ permutations of the wavevectors $\mathcal{P}(\mathbf{k}_1,\dots,\mathbf{k}_N)$.

For example, the full expression for $T^{(4)}$ is given below and corresponds to the two diagrams in Fig.\,\ref{fig:T4diag}, where we didn't label the vertices since the sum over permutations is implied
\begin{equation}
    T^{(4)}= 2^{2} A^{3}f_1^2 f_2^{2}\frac{1}{2}\sum_{ \mathcal{P}(\mathbf{k}_1,\dots,\mathbf{k}_N)}\frac{1}{k_1^3 k_2^3|\mathbf{k}_1 + \mathbf{k}_2|^3}+ A^3 f_1^3 f_3 3! \frac{1}{3!}\sum_{ \mathcal{P}(\mathbf{k}_1,\dots,\mathbf{k}_N)}\frac{1}{k_1^3 k_2^3 k_4^3}\,.
\end{equation}

It is convenient to introduce the following notation
\begin{equation}\label{eq:notation_TN}
T^{(N)}(\mathbf{k}_1,\dots,\mathbf{k}_N) = \sum_{i \in \left\{\substack{\textrm{topologies}\\\textrm{with $N$} \\ \textrm{vertices} }\right\}}T^{(N)}_{i}(\mathbf{k}_1,\dots,\mathbf{k}_N) = \sum_{i \in \left\{\substack{\textrm{topologies}\\\textrm{with $N$} \\ \textrm{vertices} }\right\}} \sum_{ \mathcal{P}(\mathbf{k}_1,\dots,\mathbf{k}_N)} t^{(N)}_{ i}(\mathbf{k}_1,\dots,\mathbf{k}_N)\,,
\end{equation}
where we split each tree-level $N$-point function into a sum of terms from different topologies each of which is further expressed as  a sum over the $N!$ permutations [$\mathcal{P}(\mathbf{k}_1,\dots,\mathbf{k}_N)$] of the external wavevectors.

Most inflationary models predict almost scale invariant $\delta_\zeta$ correlators (with conformal weight $\Delta=0$), some of which are also conformally invariant \cite{Kehagias:2013xga}.

In Primordial local Non-Gaussianity, $\delta_\zeta$ is a function of powers of $\phi_G$ only (and not its derivatives). Being that the two point function of the Gaussian field $\phi_G$ is scale invariant, we expect $\delta_\zeta$ correlators to be scale invariant as well. In the following, we demonstrate that the conformal invariance of any tree-level $N$-point correlator of $\delta_\zeta$ follows from the conformal invariance of the $N=2$ correlator of $\phi_G$. Moreover, we explicitly show that this holds separately for each topology and each permutation. Conformal invariance of the correlators implies that the conformal Ward identities are satisfied,
\begin{equation}
    \left[\left( \sum_{\substack{j=1\\j\neq\alpha}}^{N} \mathcal{K}_j \right) t^{(N)}_{ i}(\mathbf{k}_1,\dots,\mathbf{\bar{k}}_\alpha, \dots, \mathbf{k}_N)\right]_{\mathbf{\bar{k}}_\alpha = -\sum_{\beta\neq\alpha} \mathbf{k}_{\beta}} = 0\,,
\end{equation}
where
\begin{equation}
     \mathcal{K}_j\equiv 2(\Delta-3) \nabla_j - 2(\vect{k}_j \cdot \nabla_j)\nabla_j+\vect{k}_j \nabla^2_j\,.
\end{equation}
Here, $\mathcal{K}_j$ are the generators of special conformal transformations in wavevector space and $\nabla_j \equiv \partial_{\vect{k}_j}$. In our case $\Delta$ is equal to zero.

We choose the $\alpha$'th wavevector to be the dependent one and we use the overall wavevector conservation to express that wavevector in terms of the others.
Any tree-level $(N+1)-$point diagram can be constructed from an $N-$point tree-level diagram by adding a line to one of its $N$ vertices. Labelling this vertex with the wavevector ${\bf k}_\alpha$ and choosing it to be the dependent one as before, we have the recursion relation
\begin{equation}
t^{(N+1)}_{i'}(\mathbf{k}_1,\dots,\mathbf{k}_\alpha, \dots, \mathbf{k}_{N+1})=\frac{f_{m(\alpha)+1}}{f_{m(\alpha)}} t^{(N)}_{i}(\mathbf{k}_1,\dots,\mathbf{k}_\alpha, \dots, \mathbf{k}_{N})  \frac{A}{|\mathbf{k}_{N+1}|^3}
\end{equation}
where $m(\alpha)$ is the number of legs emerging from the $\alpha$'th vertex in $t^{(N)}_{i}$.
Therefore, the conformal Ward identities for the $t^{(N+1)}_{i'}$ diagram are
\begin{eqnarray}
 &&   \left[\left( \sum_{j=1, \,j\neq\alpha}^{N+1} \mathcal{K}_j \right) t^{(N+1)}_{i'}(\mathbf{k}_1,\dots,\mathbf{\bar{k}}_\alpha, \dots, \mathbf{k}_{N+1})\right]_{\mathbf{\bar{k}}_\alpha = -\sum_{\beta\neq\alpha} \mathbf{k}_{\beta}}\nonumber \\
&=&  \frac{A}{|\mathbf{k}_{N+1}|^3}  \left[\frac{f_{m(\alpha)+1}}{f_{m(\alpha)}}\left( \sum_{j=1, \,j\neq\alpha}^{N} \mathcal{K}_j \right) t^{(N)}_{i}(\mathbf{k}_1,\dots,\mathbf{\bar{k}}_\alpha, \dots, \mathbf{k}_{N})\right]_{\mathbf{\bar{k}}_\alpha = -\sum_{\beta\neq\alpha} \mathbf{k}_{\beta}} +\nonumber\\
&+&  \mathcal{K}_{N+1}\left(\frac{A}{|\mathbf{k}_{N+1}|^3}\right) \left[\frac{f_{m(\alpha)+1}}{f_{m(\alpha)}} t^{(N)}_{i}(\mathbf{k}_1,\dots,\mathbf{\bar{k}}_\alpha, \dots, \mathbf{k}_{N})\right]_{\mathbf{\bar{k}}_\alpha = -\sum_{\beta\neq\alpha} \mathbf{k}_{\beta}}\,.
\end{eqnarray}
 Using the above equation recursively the conformal invariance of the $N-$point function follows from the conformal invariance of the two-point function. The above proof can be trivially extended to $\Delta \neq 0$.

%%%%%%%%%%%%%%%%%%%%%%%%%%%%%%%%%%%%%%%%%%%%%%%%%%%%%%%%
\section{Feynman Rules for the Galaxy Bias expansion}
\label{appendix:feyn_rules}
Here we describe the diagrammatic formalism we use to graphically keep track of the different contributions in the renormalization of the operator $\delta_\rho^2$. We find the diagrammatic notation introduced here closer to the Feynman diagrams typically used in particle physics compared to what is present in the literature.

\begin{figure}[t]
\centering
\input{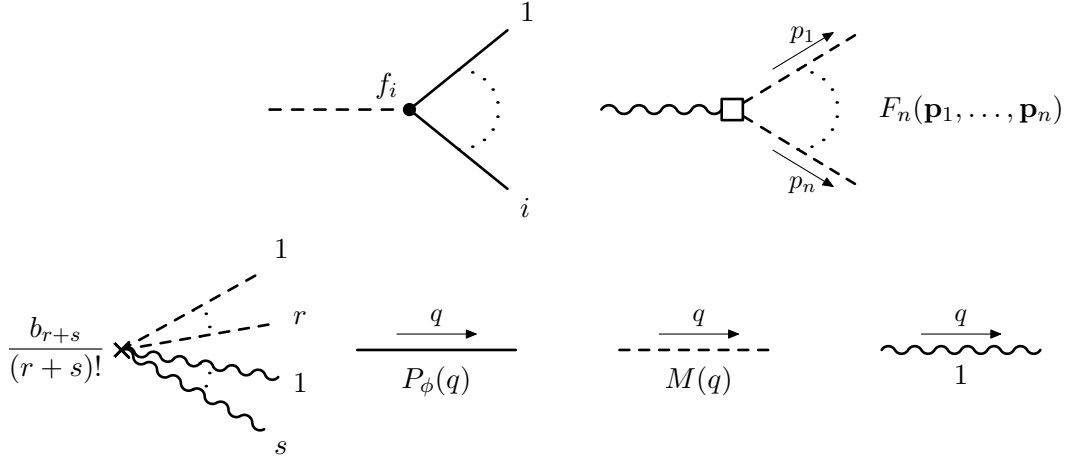}
\caption{Feynman rules for the vertices and the lines used in the computation of galaxy overdensity field correlators.}
\label{fig:feynrules}
\end{figure}

For the calculations performed in this paper we need three kinds of vertices - the dot, the box and the cross - and three kinds of lines - the solid, the dashed and the wavy. Fig~\ref{fig:feynrules} shows the factors associated with each of the above elements. The $f_i$ and $F_n$ vertices describe the effect of primordial non-Gaussianities and non-linear gravitational evolution respectively. The $b_i$ vertices represent the insertion of the operator $\left(\delta_\rho\right)^i$ in correlators of the galaxy overdensity field.
Solid and dashed lines denote the power spectrum $P_\phi$ of the Gaussian field and the transfer function $M$ introduced in Eqs.~\eqref{eq:phipropagator}~and~\eqref{eq:M}, respectively.

Valid diagrams representing contributions to the connected correlators of the galaxy overdensity field $\delta_g$ in wavevector space are drawn using the following rules,
\begin{itemize}
    \item Every $f_i$ vertex (dot) is connected to a dashed line and to $i$ solid lines;
    \item Every $F_n$ vertex (box) is connected to a wavy line and to $n$ dashed lines. Each dashed line is connected to a solid line;
    \item Every $b_i$ vertex (cross) is connected to $r$ wavy lines and $s$ dashed lines where $r+s =i$, $r,s \geq 0$.
\end{itemize} 
Diagrams with $N$ cross vertices represent contributions to the galaxy overdensity field correlator $\langle \tilde\delta_g({\bf k}_1) \dots\tilde\delta_g({\bf k}_N)\rangle$. To translate a diagram into a formula, we label the $N$ cross vertices (arbitrarily) with wavevectors $\mathbf{k}_1,\dots,\mathbf{k}_N$ conventionally considered to be incoming. Then, we multiply the factors obtained using the following rules and sum over all the distinct ($N!$) labelings of the $\mathbf{k}_i$'s.
\begin{itemize}
    \item Label each internal line with a different wavevector $\mathbf{q}_j$ and assign a factor $P_\phi(q_j)$ to solid lines, $M(q_j)$ to dashed lines and $1$ to wavy lines;
    \item Assign factors of $f_i, F_n(\mathbf{p}_1,\dots, \mathbf{p}_n)$ and $b_{r+s}/(r+s)!$ to the dot, the box and the cross vertices as shown in Fig.~\ref{fig:feynrules};
   \item 
   Further assign a factor of $(2\pi)^3\delta^3(\mathbf{k}_i-\mathbf{p})$ to each cross labelled with wavevector $\mathbf{k}_i$, and a factor of  $(2\pi)^3\delta^3(\mathbf{p})$ to each dot and box vertex, being $\mathbf{p}$ the sum of the outgoing wavevectors at the vertex. This imposes wavevector conservation at each vertex;
    \item Integrate over all the internal wavevectors $\int{d^3 {\bf q}_j}/{(2\pi)^3}$. For each loop, this procedure will leave an unconstrained wavevector ${\bf q}_i$ to be integrated over;
    \item Assign a factor that takes into account the number of possible ``Wick contractions''.
\end{itemize}
As anticipated in Sec.\ref{sec:galaxy_correlators} for loop diagrams regularization is needed to make the results finite.
The three regularization prescriptions discussed in the main text are implemented in this diagrammatic notation as follows. 
\begin{itemize}
    \item For the first regularization scheme adopted in this paper, we assign a $\theta(\Lambda-p)$ to each dashed line in a loop with wavevector $\bf p$ flowing into it;
    \item For the second regularization scheme, we assign a $\theta(\Lambda-p)$ to each line stemming from the cross vertex in the loop with wavevector $\bf p$ flowing into it;
    \item For the third regularization scheme (i.e. cutting off the convolution integral of $\delta_\rho^2$) the additional Feynman rule for loops prescribes summing over two thetas as in Eq.~\eqref{eq:regdelta2} and dividing by a factor of two. This corresponds to the two ways of \textit{routing} the loop wavevector ${\bf q}$ (e.g. Fig.\,\ref{fig:Gauss_F2}). 
\end{itemize}

We will now present an illustrative example to familiarize the reader with the notation.
\begin{figure}[t]
\input{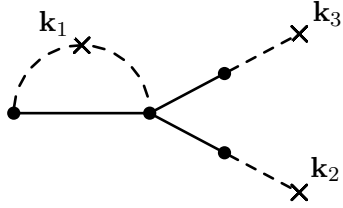}
\caption{Diagram contributing to the three point function of the galaxy overdensity field.}
\label{fig:bispectrum_example}
\end{figure}

The contribution to the bispectrum of the diagram in Fig.~\ref{fig:bispectrum_example} in the third regularization scheme is
\begin{equation}
\label{eq:bispectrum_example}
    12\, b_1^2\frac{b_2}{2} f_3 M(k_2) M(k_3)P_\phi(k_2) P_\phi(k_3)\int\frac{d^3 q}{(2\pi)^3} M(q) M(|{\bf k}_1-{\bf q}|)P_\phi(q)\,\left[\frac{\theta(\Lambda-q)+\theta(\Lambda-|{\bf k-q}|)}{2}\right].
\end{equation}
Using Eq.~\eqref{eq:M_theta} the finite parts of loop integrals can easily be estimated in the region $k_i < q_0$. For example for the integral in Eq.~\eqref{eq:bispectrum_example} we have
\begin{equation}
\int\frac{d^3 q}{(2\pi)^3} P_\phi(q)M(q)\left[ M(|{\bf q}-{\bf k}_1|) -M(q)\right] \theta(\Lambda-q)\, \isEquivTo{|{\bf k}_1| < q_0} \frac{A}{2\pi^2}\left(\frac{2}{5\Omega_m}\right)^2\frac{k_1^2 q_0^2}{H_0^4}\,.
\end{equation}

\begin{figure}[t]
\centering
\input{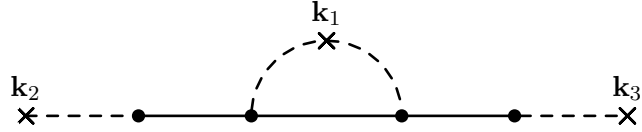}
\vspace{-1cm}
\caption{Example of an infrared divergent diagram.}
\label{fig:example_IR}
\end{figure}

In the text we addressed infrared divergences by expanding the integrands of loop diagrams and truncating the infrared divergent parts. Alternatively, infrared divergences in correlators of the galaxy overdensity field can be removed by introducing an infrared regulator $m$, modifying the tree-level power spectrum of the primordial curvature fluctuations as 
\begin{equation}\label{PS_IR_safe}
P_{\phi}(q)\to \frac{A}{(q^2+m^2)^{3/2}}\,.
\end{equation}
For example, the contribution to the galaxy overdensity bispectrum in Fig.~\ref{fig:example_IR} is
\begin{equation}
2 \frac{b_2}{2} b_1^2 f_2^2  M(k_2) M(k_3)P_{\phi}(k_2)P_{\phi}(k_3) \sigma^2({\bf k}_1, {\bf k}_2; \Lambda,m)+\{{\bf k}_2 \leftrightarrow {\bf k}_3 \}\,
\end{equation}
where
\begin{equation}
    \sigma^2({\bf k}_1, {\bf k}_2; \Lambda,m)=\int\frac{d^3 q}{(2\pi)^3} \frac{A M(q) M(|{\bf k}_1 - {\bf q}|)}{(|{\bf q}+{\bf k}_2|^2+m^2)^{3/2}}\theta(\Lambda-q)\,.
\end{equation}
Most of the dependence on the infrared regulator $m$ in the above integral is from the region of integration where the argument of $M$ is less than $q_0$. We find that
\begin{equation}
    \sigma^2({\bf k}_1, {\bf k}_2; m;\Lambda) = \left(\frac{2}{5 \Omega_m H_0}\right)^2 A\frac{k_3^2|{\bf k}_1-{\bf k}_2|^2}{4\pi^2}\log{\left(\frac{q_0^2}{m^2}\right)}+ \textrm{IR finite as } \frac{m}{q_0} \to 0\,.
\end{equation}
For a wide range of $m/H_0\ll 1$, the correlators of the galaxy overdensity field depend very weakly on the exact value of $m$.

\end{document}